\documentclass[11pt,letter]{article}
\usepackage{graphicx}
\usepackage{amsmath}
\usepackage{amssymb}
\usepackage{cancel}
\usepackage[left=2cm,right=2cm,top=3cm,bottom=2cm]{geometry}
\usepackage{setspace}
\usepackage{float}
\usepackage[super,sort&compress,comma]{natbib}
\usepackage[font={footnotesize}]{caption}
\usepackage[labelfont=bf]{caption}
\usepackage{bm}
\usepackage[T1]{fontenc}
\usepackage{microtype}
\usepackage{subfigure}
\usepackage{chemformula}
\usepackage[version=4]{mhchem}
\usepackage{textcomp}

\usepackage{color}
\include{MyCommand}

\usepackage{tabularx}
\usepackage{booktabs}
\usepackage{multirow}
\usepackage{mathtools}
\usepackage{bm}
\usepackage{gensymb}
\usepackage{esvect}

\begin{document}
\title{\textbf{The Interplay of Pauli Repulsion, Electrostatics, and Field Inhomogeneity for Blueshifting and Redshifting Vibrational Probe Molecules}}
\date{}
\author{R. Allen LaCour*$^{1,2}$, Ruoqi Zhao$^{1,2}$, Teresa Head-Gordon$^{*1,2,3}$}
\maketitle
\noindent
\begin{center}
$^1$Kenneth S. Pitzer Theory Center and Department of Chemistry\\
$^2$Chemical Sciences Division, Lawrence Berkeley National Laboratory\\
$^3$Departments of Bioengineering and Chemical and Biomolecular Engineering\\
University of California, Berkeley, CA, 94720 USA

corresponding author: alacour@lbl.gov,thg@berkeley.edu
\end{center}

\begin{abstract}
\noindent
Many molecules' vibrational frequencies are sensitive to intermolecular electric fields, enabling them to probe the field in complex molecular environments. However, it is often unclear whether the probe is responding to the local electric field or other types of intermolecular interactions, inhibiting interpretation of the frequency and effectiveness as probes. This is especially true of molecules whose vibrational frequencies blueshift instead of the more typical redshift in hydrogen bonding configurations. Here we computationally investigate the causes of redshifting versus blueshifting over a range of vibrational reporters. First, we apply adiabatic energy decomposition analysis to a paradigmatic set of probes, finding that redshifting only occurs when electrostatic interactions are strong enough to overcome the dominant and large blueshifting contribution of Pauli repulsion. Furthermore, we demonstrate that field inhomogeneity can further shift the frequency of many probes substantially to either reinforce or counteract the shift expected from a homogeneous field. We find that redshifting is reinforced by electric field inhomogeneity, otherwise field inhomogeneity further weakens the electrostatic contribution relative to Pauli repulsion, leading to blueshifting. Further calculations indicate that the probe's response to field inhomogeneity can be understood by considering the mass of the atoms involved in the stretching mode and sign of the electric field. In explaining the interplay of different intermolecular interactions and field inhomogeneity for many probes, our results should enable the use and interpretation of spectroscopic probes and their connection to electric fields in more complex systems.

\end{abstract}

\section{Introduction}

Electrostatic interactions are important driving forces in complex chemical processes\cite{Boxer2025} like catalysis and  reactivity\cite{warshel1998electrostatic,  fried2014extreme, ma2015site,Welborn2018,leonard2021electric}, aqueous solvation and pH\cite{chandler2005interfaces,Marenich2008,Case2019}, and chemical and conformational equilibria\cite{Schaefer1997, hunter2012nature,zhou2018electrostatic,Judd2025,Tsanai2025}. Unsurprisingly then many spectroscopic methods have been developed to characterize electrostatic interactions, often in the form of electric fields.\cite{fried2013measuring,fried2015measuring,Xiong2020,Lake2024,shi2025water}
Vibrational spectroscopic probes, which possess vibrational frequencies that respond to their environment in a known manner, are one flexible measure of the electrostatic intermolecular interaction.\cite{kim2013infrared, feng2024unnatural, blasiak2017vibrational, baiz2020vibrational}.
Common probes include added small molecules, like the CO stretch of carbonyls\cite{park2002origins, fried2013measuring} or the CN stretch of nitriles\cite{andrews2000vibrational, lindquist2009nitrile}, or \textit{in situ} probes such the amide groups of proteins\cite{lin2009empirical, bondarenko2015application, maekawa2010comparative} and the OH stretch of water\cite{fecko2003ultrafast, corcelli2004combined, bakker2010vibrational, lacour2023predicting, shi2025water}.
Further possibilities are afforded by isotopic labels\cite{mirkin2004new, smith2005unified} like $^2$D or $^{13}$C, which can shift the frequency to a less congested part of the spectrum or decouple vibrational modes altogether.
Numerous systems, including interfaces\cite{stiopkin2011hydrogen, pullanchery2021charge, shi2025water, parsons2026s} and proteins\cite{tadesse1991isotopically, lin2009empirical, reppert2016computational, lorenz2020infrared}, have been interrogated with vibrational spectroscopic probes.

Most spectroscopic vibrational probes consist of a stretching mode with a terminal atom exposed to the intermolecular environment. Its vibrational frequency will respond to an electric field if the molecule's multipole moments change during the vibration. If the stretching mode is decoupled from other vibrational modes and charge polarization is minimal, the influence of the electric field on the mode's potential energy surface can be written as:
\begin{equation}
    \Delta U(R) = -\boldsymbol{E}\cdot \boldsymbol{\mu}(R) - 1/2\cdot\boldsymbol{\nabla} \boldsymbol{E}:\boldsymbol{Q}(R) \space...
\label{eq:1}
\end{equation}
where $R$ is the bond length of the stretching mode, $\boldsymbol{E}$ is any environmental (non-intramolecular) electric field, $\boldsymbol{\mu}$ is the dipole, and $\boldsymbol{Q}$ is the quadrupole. If the magnitude of $\boldsymbol{E}$ is much larger than any of its gradients, Equation \ref{eq:1} can be truncated at the dipole term, and the probe's ability to report on electric fields stems entirely from how $\boldsymbol{\mu}$ depends on the bond length. Aligning $\boldsymbol{\mu}$ with the electric field will elongate the bond and is more energetically favorable than anti-aligned configurations which contract the bond. Due to bond anharmonicity, bond elongation or contraction results in redshifting or blueshifting, respectively, by an amount dependent on the magnitude of the field. The relationship between field strength and vibrational response underlies the vibrational Stark shift experiment that has been used to characterize electric field environments in proteins\cite{fried2013measuring,li2021} and at  interfaces\cite{Xiong2020,sarkar2021advances,shi2025water}.

The effective use of spectroscopic probes to monitor electric fields depends on how well their frequencies can be related to their local environment. Often shifts in the frequency are attributed to the electric fields generated by neighboring molecules (``intermolecular electric fields'')\cite{bublitz1997stark, boxer2009stark,fried2015measuring, sarkar2021advances}, but it has recently become clear that many molecules appear sensitive to other types of intermolecular interactions.
This appears especially true of several stretching modes that blueshift in hydrogen bonding environments, such as the CH stretches in ``blueshifting hydrogen bonds''\cite{hobza2000blue, li2002physical, hermansson2002blue, mo2014nature} and several probes with nitrogen-containing groups.\cite{andrews2000vibrational, fafarman2010decomposition, maj2016isonitrile, you2019isonitrile}. Instances of blueshifting have proven difficult to understand, with many putative explanations put forth.
Several recent studies have implicated Pauli repulsion\cite{blasiak2016vibrational, wang2017unified,xu2018direct, mao2019probing, zhao2022origin} which may induce blueshifting from the ``mechanical compression''\cite{mao2019probing} that other molecules can exert on a stretching mode. Another source arises from spatial variation or "inhomogeneity" in the electric field\cite{lee2012vibrational,choi2013computational}; all intermolecular electric fields are inhomogeneous, and thus truncating Eq. \ref{eq:1} at the dipole level is only ever approximate, thus 
field inhomogeneity has been linked to blueshifting through interaction with the molecular quadrupole\cite{lee2012vibrational, kirsh2024hydrogen}. These important insights, however, have not completely revealed which of these factors are significant or how they work together to create blueshifting or redshifting probes, limiting our ability to construct general spectroscopic maps and to aid interpretability.

Here, we computationally investigate blueshifting and redshifting in the vibrational frequencies of several probe molecules to better understand their origin and to aid in their interpretative value. First, we demonstrate using energy decomposition analysis (EDA) that Pauli repulsion and permanent electrostatics are the most substantial contributions to vibrational frequency trends.
While Pauli repulsion provides a consistently large blueshift for all reporters, the electrostatic contribution tends to be weak in blueshifting probes but strongly redshifting in redshifting probes. We can explain this variation by considering how probes respond to electric field inhomogeneity. 
In blueshifting probes, the field inhomogeneity counteracts the shift expected for a homogeneous electric field, resulting in an electrostatic contribution significantly weaker than Pauli repulsion, whereas red shifting probes reinforce the shift expected for a homogeneous electric field thereby overcoming Pauli repulsion. We also find that we can anticipate the behavior of different probes considering the mobility of the atoms involved in the vibration and the sign of the charge generating the electric field. By revealing the interplay between different intermolecular interactions and the role of field inhomogeneity, our work provides insight for the future development and interpretation of spectroscopic reporters to interpret frequency shifts in complicated media.

\section{Methods}

We use EDA calculations (Figure 2) to probe the vibrational frequencies of 6 probes:
HOH, HCCH, (CH$_3$)$_2$CO, F$_3$CH, CH$_3$CN, and CH$_3$NC.
When examining the response of probes to electric fields (Figure 3), we examine the same probes with the addition of FH and C$_6$H$_6$.
Finally, when examining homonuclear diatomics (Figure 4), we examine HH, FF, OO, and NN.
In every case, the stretching frequency of interest is given by the last two atoms listed for the molecule.
Rather than use the full normal mode, we make the approximation of just examining the frequency produced by two primary atoms involved in the stretch.
Despite this approximation, we correctly reproduce the redshifting and blueshifting of different probes, indicating that our focus on the stretching coordinate captures much of the physics involved in frequency shifting.

All EDA calculations were performed with the Q-Chem 6.1 software package\cite{epifanovsky2021software} at the $\omega$B97X-V/aug-cc-pVTZ level of theory\cite{mardirossian2014omegab97x}.
We used 99 radial and 590 angular grid points for the exchange-correlation function and 50 radial and 194 angular gridpoints for the VV10 correction\cite{vydrov2010nonlocal}.
We used the energy decomposition analysis (EDA) based on the absolutely localized molecular-orbitals (ALMO) scheme\cite{khaliullin2007unravelling, horn2016probing} for decomposing the energy (and forces) into the following terms:
\begin{equation}
 E_\mathrm{int}\ = E_\mathrm{elec}\ +\ E_\mathrm{Pauli}\ +\ E_\mathrm{disp}\ +\ E_\mathrm{pol}\ + E_\mathrm{CT}\   
\end{equation}        The quantities $E_\mathrm{elec}$ and $E_\mathrm{Pauli}$ are predominately 2-body terms describing permanent electrostatics and Pauli repulsion, while $E_\mathrm{disp}$, $E_\mathrm{pol}$, and $E_\mathrm{CT}$ correspond to the different contributions from dispersion, polarization, and charge transfer, respectively.

The EDA calculations were performed in two parts using the water molecule to examine hydrogen bonding with multiple vibrational probes. First, for a given distance from the probe to the H$_2$O, we optimized all degrees of freedom except for the distance between the nearest atom of the probe and the atoms involved in the stretching mode.
With one exception, we kept the angle between the atoms in the stretching mode and nearest atom of the probe molecule linear.
The exception was (CH$_3$)$_2$CO, for which we allowed the angle to vary.
Next, we held the positions of all other atoms constant while varying the distance between the two atoms in the stretching mode.
We varied their distance by moving them by an amount inversely proportional to their mass, and thus their center of mass did not change. 
We examined 16 such stretch distances within roughly 0.3 \AA\space of the equilibrium distance, performing EDA for each of them.
We then fit a sixth order polynomial to each EDA component and the overall energy as a function of stretch distance.
Using the polynomial fits, the force from each component was computed by taking its derivative with respect to stretch distance at the minimum energy-stretch distance.

To investigate the influence of field inhomogeneity, we examine the response of several spectroscopic probes to the electric field generated by a point charge, which we illustrate for aligned configurations for water and acetonitrile in Figures 1a and 1b, respectively. As shown in Figure 1c for acetonitrile, the field generated by a point charge becomes more inhomogeneous as the point charge approaches the molecule. To isolate the role of field inhomogeneity, we vary the charge of the point charge to keep the magnitude of the field at the midpoint of the stretch fixed while moving the point charge closer to the probe. We kept every atom of the probe molecule fixed except for the two atoms involved in the stretching mode. We set the origin to be the midpoint between these two atoms, while the angle between the charge and the two atoms involved in the stretching mode was kept linear, as illustrated in Figures 1a and 1b, with the exception of (CH$_3$)$_2$CO in which the angle between the stretching mode and the charge was 120\degree.

To get the frequency, we performed individual single point energy calculations for different stretch distances. We varied the stretch distance by moving the atoms further or closer to the origin by an amount inversely proportional to their mass.
For each value of the field and position of the point charge, we examined 16 stretch distances.
We then fit the energy at each stretch distances to a sixth degree polynomial. We obtained the frequency by applying the harmonic approximation to the minimum of the polynomial. The calculations for a point dipole were performed similarly, except instead of a single point charge we used two of equal and opposite signs. These were placed 0.1 \AA\space apart from each other in a linear arrangement with the stretching mode. The sole exception is again (CH$_3$)$_2$CO, for which the angle between the stretching mode and the two charges was 120\degree. For every probe at  $D = \infty$, we used a homogeneous field rather than a point charge in our calculations.

\begin{figure}[H]
\centering
\includegraphics[width=0.75\textwidth]{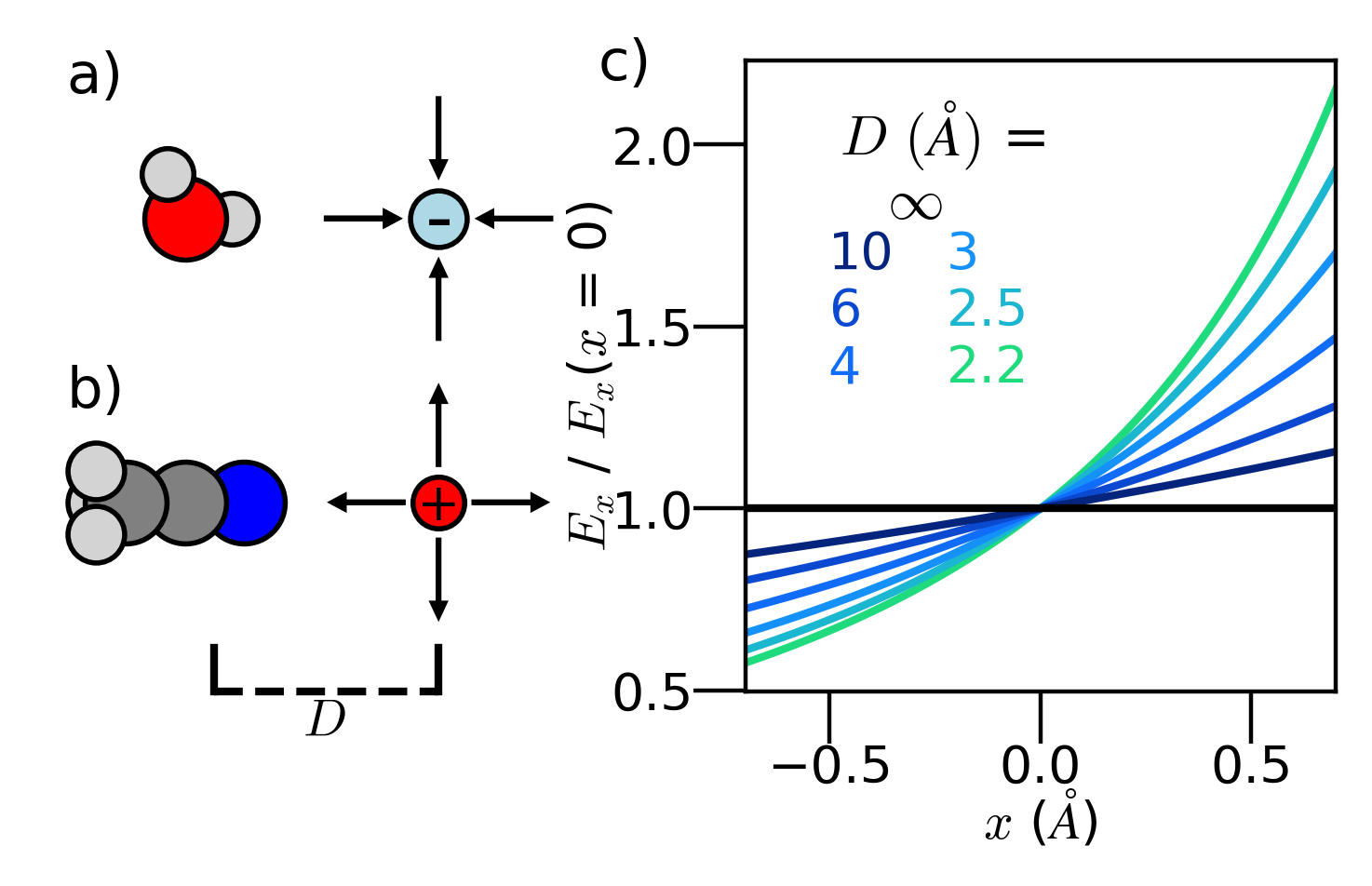}
\caption{\textbf{The electric field produced by point charges.} Illustration of the interaction of (a) HOH and (b) CH$_3$CN with point charges. Both molecules prefer to align their dipoles with the field. For HOH, this corresponds to the H pointing towards a negative charge. For CH$_3$CN, this corresponds to the N pointing towards a positive charge. In (b) we also illustrate the coordinate system used throughout the text, in which the midpoint of the probe stretch is at $x=0$ and the point charge is $D$ distance away. In (c) we show how the field experienced by the probe depends upon the position of a positive point charge.
We specifically show the $x$-components of the field (the other components are zero) relative to a homogenous field with same magnitude at the bond midpoint. As the point charge approaches the probe, the spatial variation, or inhomogeneity, of the field increases.}
\label{fig:fig1}
\end{figure}

To get the anharmonic frequencies shown in Figure S3, we performed variational calculations in a manner similar to our previous work\cite{lacour2023predicting}, except we only considered the one-dimensional energy surface of the stretch here.
We used the same energy surfaces computed for our harmonic calculations in our variational calculations.
Our basis functions $\psi$ were the 16 lowest eigenfunctions of the one-dimensional harmonic oscillator.
We evaluated the Hamiltonian matrix $H_{ij} = \langle\psi_i \vert H \vert \psi_j \rangle$ numerically and diagonalized it to obtain the eigenvalues.
We computed the frequency as the difference between the ground- and first-excited-state eigenvalues.

\section{Results}

\subsection{Energy Decomposition Analysis}
We begin by using the absolutely localized molecular orbital EDA (ALMO-EDA)\cite{khaliullin2007unravelling, horn2016probing} to measure the contribution of different intermolecular interactions to the stretching frequencies of six probe molecules as reporters for frequency shifts in different chemical environments. As probes we selected three molecules known to redshift in hydrogen-bonding (H-bond) configurations\cite{zheng2025beyond} (HOH, HCCH, and (CH$_3$)$_2$CO) and three molecules known to blueshift in H-bond configurations\cite{hobza2000blue, maj2016isonitrile} (F$_3$CH, CH$_3$CN, CH$_3$NC). We examined how their probe frequency varied in response to interaction with a water molecule, which corresponds to a typical H-bond environment.  Depending upon which atom the probe forms hydrogen bonds with, we placed the O or the H of the water closer to the terminal atom of the stretching mode, as illustrated in Figure 2. We then optimized the position of every atom in the probe and the water except for the two atoms involved in the stretching mode and the nearest atom of the water. Except for (CH$_3$)$_2$CO, which prefers a smaller angle, we kept the angle between these three atoms linear during the optimization. We did this for different distances between the nearest atom of the water and the terminal atom of the stretch. We denoted this distance relative to the minimum-energy distance between the probe and the water as $\Delta$$d_{min}$. We then decomposed the force on the probe stretch into different contributions; see Methods for complete details on the calculations.

\begin{figure}[H]
\centering
\includegraphics[width=0.85\textwidth]{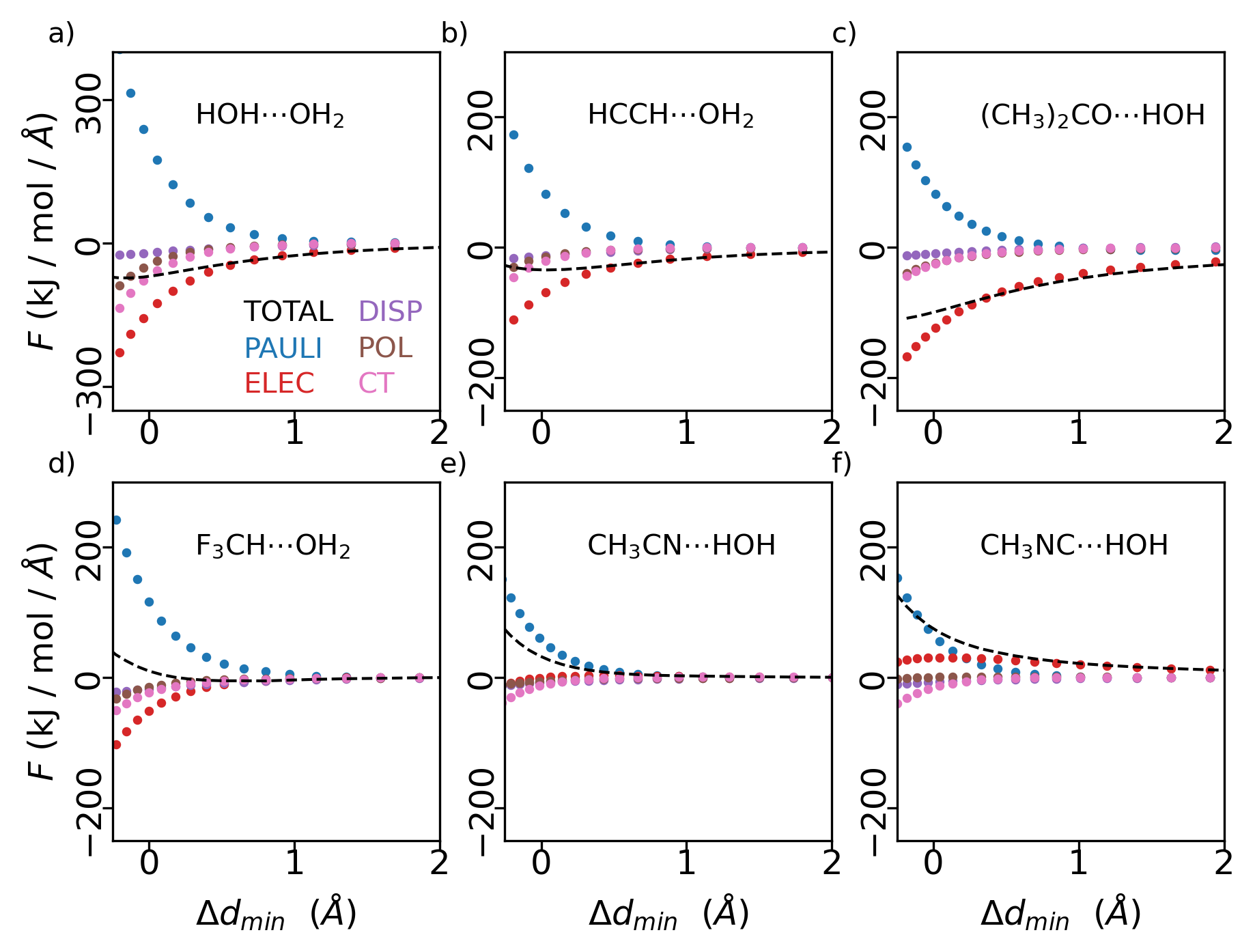}
\caption{\textbf{Energy decomposition analysis for probes interacting with water.} In each panel, we use EDA to analyze the different contributions to the probe frequencies. The dashed line marked "Total" indicates the overall force on the stretching mode, while other lines represent the contributions from specific intermolecular interactions. Positive forces contract the bond and induce blueshifting, while negative forces expand it to induce redshifting. $\Delta d_{min}$ is the distance between the terminal atom of the probe's stretching mode and nearest molecule of the water relative to their distance at the energy minimum. The probes  (a) HOH, (b) HCCH, and (c) (CH$_3$)$_2$CO all redshift\cite{zheng2025beyond} in H-bond configurations. The probes (d) F$_3$CH, (e)  CH$_3$CN, (f) CH$_3$NC all blueshift\cite{hobza2000blue, maj2016isonitrile} in H-bond configurations. 
}
\label{fig:fig2}
\end{figure}

The results in Figure \ref{fig:fig2} show that positive forces induce bond contraction and, due to bond anharmonicity, blueshifting, whereas negative forces induce bond extension and redshifting.
We show both the total force on the stretching mode and its decomposition into five intermolecular components: Pauli repulsion (PAULI), electrostatics (ELEC), dispersion (DISP), polarization (POL), and charge transfer (CT). The probes in Figures 2a-c all redshift in H-bond environments, which is consistent with the total force being negative. Likewise, the probes all blueshift in H-bond environments in Figures 2d-f, which is consistent with the total force being positive. In all cases, Pauli repulsion is the largest blueshifting contribution near the minimum-energy distance, and is consistent with results reported previously using ALMO-EDA\cite{mao2019probing, blasiak2016vibrational}, but here we find it for a broader swathe of vibrational reporters. We note that, although its sign is always positive, the magnitude of the Pauli repulsion contribution is larger for probes with a terminal H. This reflects the fact that the small mass of H makes it highly mobile and more likely to penetrate closer to another molecule during a vibration.

Another observation is that a probe can only redshift when other intermolecular forces exceed the dominance of Pauli repulsion. Of the remaining forces, electrostatics is frequently the strongest competing interaction to Pauli repulsion near the equilibrium distance, and is hence why vibrational reporters are used to understand electric fields. Given the consistency in the behavior of Pauli repulsion regardless of probe, these results indicate that we can largely understand whether a probe blueshifts or not by considering the strength and direction of the electrostatics contribution. Of course, other interactions, like polarization and charge transfer, also play a role, but they are always much weaker than Pauli repulsion and electrostatics.

\subsection{Inhomogeneous Field from a Point Charge}
All intermolecular electric fields are inhomogeneous, and thus truncating Eq. \ref{eq:1} at the dipole level is only ever approximate. While ALMO-EDA analysis shows that the electrostatics component plays a large role in redshifting versus blueshifting, it does not tell us why the interaction varies so strongly for different probes.
In general, the electrostatics component is determined by how the probe's charge density interacts with the electric fields generated by other molecules. These intermolecular fields may shift the frequency if the probe's charge density changes during the vibration. The intermolecular fields in hydrogen bonding environments are in general highly inhomogeneous, which has been implicated as
important for blueshifting probes\cite{lee2012vibrational, mao2019probing, kirsh2024hydrogen}. However, the relevance of field inhomogeneity has recently been called into question, in part because EDA calculations like ours indicate that Pauli repulsion is the primary contributor \cite{blasiak2016vibrational, mao2019probing}.

We next seek to understand the probe's response to electric fields in situations without other intermolecular interactions. To isolate the effects of field inhomogeneity, we examine the response of the probes to the electric field generated by a point charge, whose field inhomogeneity increases as the point charge approaches the probe. Such a model has been used previously\cite{kirsh2024hydrogen}, but here we isolate the role of field inhomogeneity by changing the magnitude of the point charge to keep the field at the stretch's midpoint constant while varying its distance to the probe.
We vary the sign of the point charge to account for whether the probe prefers to interact with atoms carrying partial negative or partial positive charges. As before, we keep the angle between the point charge and probe stretching mode linear except for (CH$_3$)$_2$CO; see Methods for complete details.
We used the same probes as in Figure 2 with an additional one that redshifts when H-bonded (FH) and an additional one that blueshifts when H-bonded (benzene).
We keep the fields small to avoid significant polarization of the probe, although some can still be seen in the nonlinearity present at larger field strengths.
From Figure 3 we see that polarization of the probe only leads to greater redshifting.

\begin{figure}[H]
\centering
\includegraphics[width=0.99\textwidth]{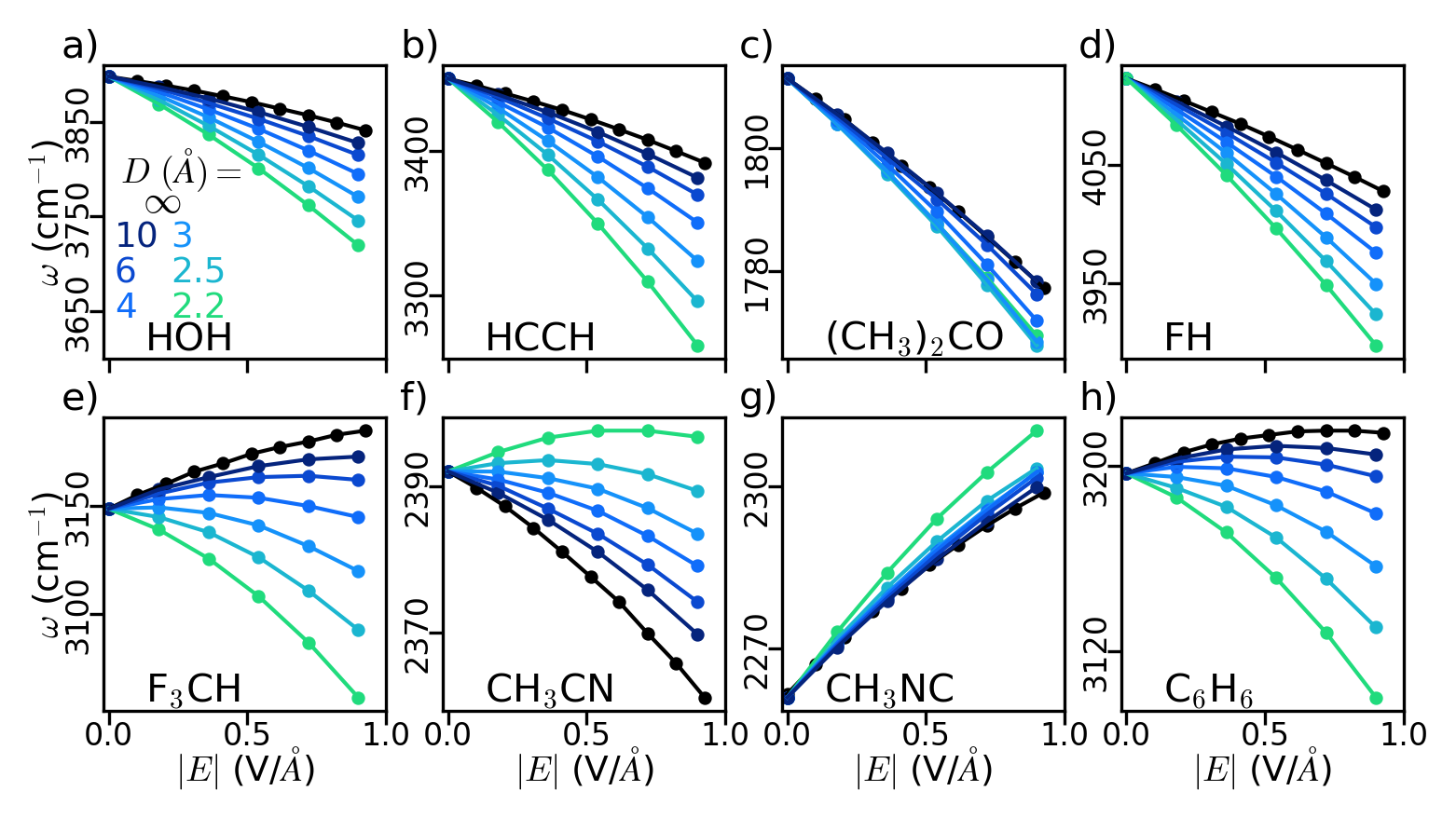}
\caption{\textbf{The frequency response of probes to the electric field generated by point charges.} Each panel shows the relationship between the stretching frequency ($\omega$) and the electric field ($\boldsymbol{E}$) experienced at the midpoint of the stretch. 
The electric field we report is just that generated by the point charge.
The sign of the point charge was chosen based upon whether the probe prefers to interact with atoms carrying partial negative (a, b, d, e, h) or partial positive (c, f, g) charges. The specific molecules examined are given in the bottom-left corner of each panel; the relevant stretching mode is between the last two atoms given. 
The distance ($D$) is the distance between the point charge and the midpoint of the bond.
Smaller $D$ indicate greater field inhomogeneity.}
\label{fig:fig3}
\end{figure}

The results in Figure \ref{fig:fig3} show some commonalities and some differences in how the probes respond to electric fields. All probes that exhibit overall redshifting in hydrogen bonding environments (a-d) are similar in that they redshift in a homogeneous field which is further reinforced due to field inhomogeneity. The degree of shifting due to field inhomogeneity does vary for different probes, with (CH$_3$)$_2$CO experiencing a weaker redshift than the other probes, but the sign of the shift is consistent. The results in Figure \ref{fig:fig3}(a-d) provide a natural explanation for the electrostatics component computed from ALMO-EDA: all have large, negative electrostatics components because the redshift in a homogeneous field is further reinforced by field inhomogeneity. 

The behavior of the overall blueshifting probes (e-h) is less consistent. F$_3$CH and benzene blueshift in a homogeneous field, but the blueshift becomes weaker and ultimately turns into a redshift as field inhomogeneity increases. Conversely, CH$_3$CN redshifts in a homogeneous field, but the redshift weakens until it turns into a blueshift under the strongest field inhomogeneity. Finally, CH$_3$NC blueshifts in a homogeneous field and further blueshifts as the field grows more inhomogeneous. These results are reconciled with ALMO-EDA by the fact that they all have weaker electrostatics components compared to the redshifting probes. The smaller electrostatics component for CH$_3$CN relative to F$_3$CH agrees with the smaller scale of the frequency shift observed for CH$_3$CN for point charges closer than 3 \AA. When both a homogeneous field and field inhomogeneity induce blueshifting, as with CH$_3$NC, the electrostatics component itself is blueshifting as seen in Figure \ref{fig:fig2}. As expected, the probe frequencies also depend upon whether their bond dipole is aligned or anti-aligned with an electric field as seen in Supplementary Figure S1. 
We also note that the equilibrium bond length changes with the electric field in a manner inverse to the frequency, as shown in Figure S2, and that the trends shown in Figure 3 are unaffected by anharmonicity, as shown in Figure S3.

Of course, the point charge model is simpler than actual intermolecular environments. Nonetheless, the observed qualitative agreement indicates that it captures the probes' response to field inhomogeneity in a manner similar to real H-bond environments. To test a different charge distribution, we also examined the response of these same probes to the fields generated by point dipoles, whose fields are significantly more inhomogeneous than those of point charges at the same distance from the probe. As shown in Supplementary Figure S4, the molecules respond similarly to the fields generated by dipoles and to those generated by point charges.
The shift due to field inhomogeneity is larger than with point charges, which is consistent with the increase in field inhomogeneity.

\subsection{Additional Features of an Inhomogeneous Field on Vibrational Shifts}
While the probe response to a homogeneous field is captured entirely by the dipole moments, the close distances relevant for hydrogen bonds limit the use of higher-order multipoles in the quantitative analysis of changes in vibrational frequencies with inhomogeneous fields. Nonetheless, we can identify two factors that capture much of the impact of electric field inhomogeneity. First, we find that the field experienced by the more mobile atom during a vibration appears more relevant for the frequency than the field at the bond midpoint. If the field experienced by the more mobile atom is more relevant, this increase in field strength could lead to a greater shift in the frequency for probes with a terminal H, which are highly mobile in stretching modes. Indeed, Figure \ref{fig:fig3} shows that all species with a terminal H experience large redshifts due to field inhomogeneity, while the effect of field inhomogeneity on stretching modes is much weaker for probes without an H. We further demonstrate this point in Supplementary Figure S5, where we find that altering the masses of atoms in the HF molecule can dramatically reduce the effects of field inhomogeneity.
This finding supports the idea of measuring the electric field at the H as opposed to the midpoint for stretches containing H, as done by Skinner et al.\cite{corcelli2004combined,schmidt2004ultrafast} for many years.

\begin{figure}[H]
\centering
\includegraphics[width=0.825\textwidth]{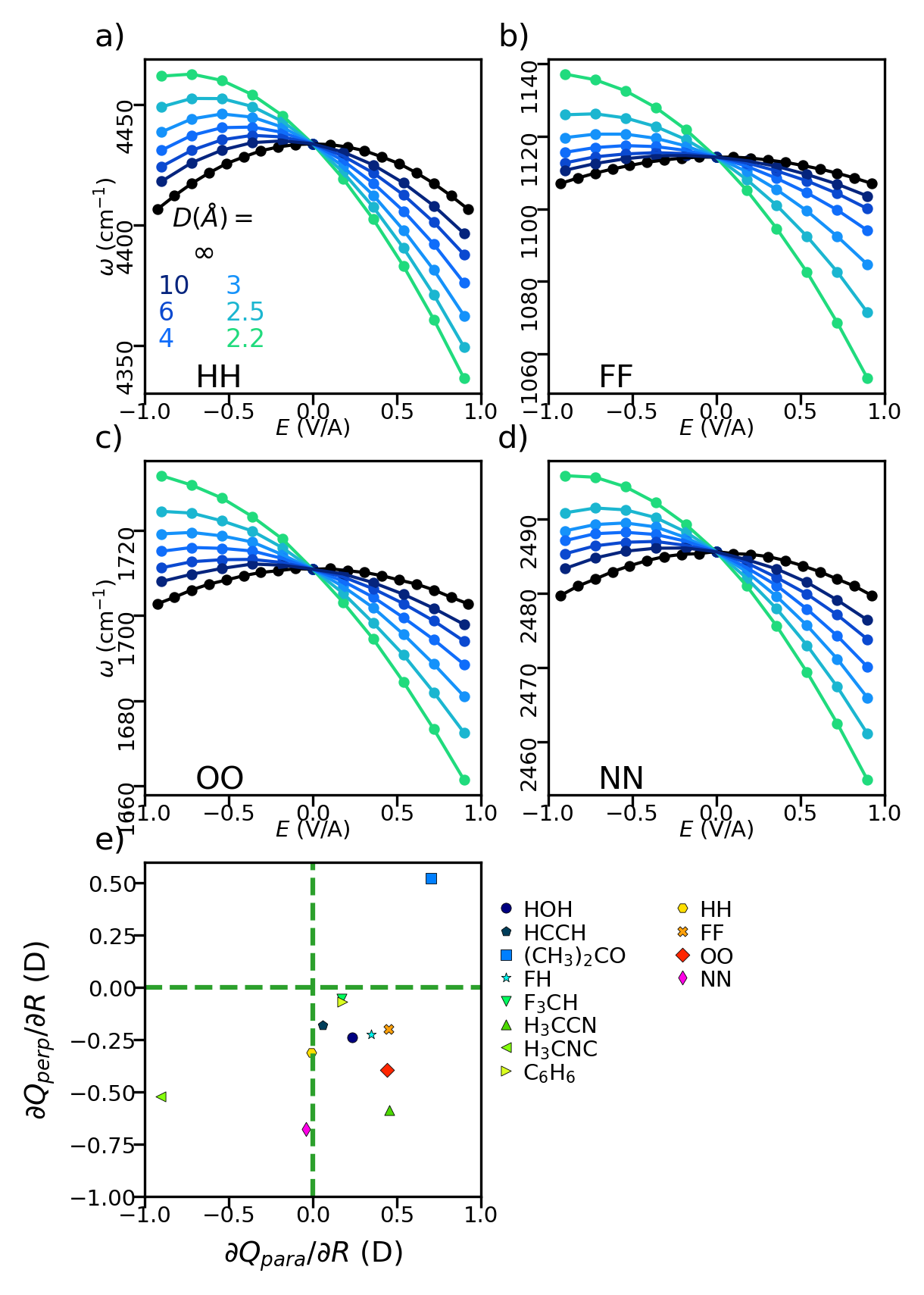}
\caption{\textbf{The response of homonuclear diatoms to field inhomogeneity.} The homonuclear probe molecules have no dipole moment, thus we report the component of the electric field in the $x$ direction without projecting it along the bond dipole. We show the frequency response of (a) HH, (b) FF, (c) OO, and (d) NN to the field generated by a point charge.
Positive and negative fields were generated by negative and positive point charges, respectively.
In (e), we examine the derivatives of the quadrupole with respect to directions parallel to and perpendicular to the bond axis for all probes examined in this study. We add the dashed lines to highlight which quadrant the values falls into.}
\label{fig:fig4}
\end{figure}

Second, we find that stretching modes respond asymmetrically to fields generated by positive or negative point charges. To demonstrate this, we investigated the stretching frequency of several homonuclear diatomic molecules in electric fields by examining their response to charges of both signs as demonstrated in Figure \ref{fig:fig4}(a-d). As is expected for symmetric molecules, the response of these molecules to a homogeneous field is completely symmetric with respect to the sign of the field. However, the symmetry is broken with increasing inhomogeneity. For all four homonuclear diatomics, a positive field induces redshifting while a negative field induces blueshifting. With one exception, this is consistent with the results in Figure \ref{fig:fig3}, in which species that interact with negative (positive) charges experience redshifting (blueshifting) due to field inhomogeneity. The one exception is (CH$_3$)$_2$CO. As before, its exceptional behavior is likely attributable to the non-linear angle at which it forms hydrogen bonds.

To understand the dependence on the sign of the point charge, we next examine the contribution of the quadrupole. Although contributions from higher order multipoles prohibit quantitatively correlating the quadrupole with the frequency shifts, the quadrupole is a measure of the variance of charge density and so may provide a qualitative correlation.
For a  probe interacting with a point charge in the geometry shown in Figure \ref{fig:fig1}, the quadrupole contribution to the energy can be written as:
\begin{equation}
    \Delta U_{Q}(R) = 
    -kq\frac{Q_{para} - Q_{perp}}{{D}^3}
\label{eq:2}
\end{equation}
where $k$ is Coulomb's constant, $q$ is the charge of the point charge, and $Q_{para}$ and  $Q_{perp}$ are the parallel and perpendicular components of the quadrupole, respectively. As with the dipole moment, higher order multipoles may shift the frequency if the multipole varies with bond length.
In Figure \ref{fig:fig4}e, we plot the derivatives of the quadrupole with respect to bond length for each probe. We take $Q_{perp}$ to be the average of the quadrupole components perpendicular to the bond if they are not equal.
The behavior seen here, in which positive charges induce blueshifts and negative charges induce redshifts, is consistent with $\partial Q_{para}/ \partial R > 0$ and $\partial Q_{perp}/\partial R < 0$. As shown in Figure \ref{fig:fig4}e, almost every molecule examined here is clustered near that quadrant. If we interpret the quadrupole moments as the variance of the electron density, the trend in Figure \ref{fig:fig4}(e) implies that, as bonds extend, the electron density typically expands perpendicularly to the bond but contracts parallel to the bond. This is apparently quite general, with the exceptions being (CH$_3$)$_2$CO and CH$_3$NC.  We note that both were outliers in Figure \ref{fig:fig3} in that they had the lowest sensitivity to field inhomogeneity, suggesting that higher order multipole moments may play a larger role for them.

\section{Discussion and Conclusions}
Here we investigated how different intermolecular interactions contribute to blueshifting versus redshifting in vibrational frequencies. Using ALMO-EDA, we found that Pauli repulsion and electrostatics were frequently the interactions which most shifted the vibrational frequencies. While Pauli repulsion contributes a consistent blueshift, the electrostatics component varies from strongly redshifting to blueshifting for different molecules.
Importantly, the electrostatics component was strongly redshifting in the overall redshifting probes because only then could the blueshift from Pauli repulsion be overcome.

We interpreted the variation in the probe's electrostatics components by considering their response to homogeneous and inhomogeneous electric fields. When field inhomogeneity reinforces the redshift expected from a homogeneous field, the electrostatic contribution is strongly redshifting.
When the shift from field inhomogeneity countered the shift from a homogeneous field (or $vice$ $versa$), the electrostatic contribution was weak, enabling Pauli repulsion to dominate. Electric fields thus influence the behavior of overall blueshifting probes even though EDA indicates that the blueshifting is primarily due to Pauli repulsion. We found that we could understand the responses of different probes by considering atomic mobility during vibrations and asymmetry in their response to electric fields of different signs.

We note that our results have implications for the usefulness of these molecules as reporters of their local environment. Most of the stretching modes we examined have been used as reporters in some regard, but our results indicate that some may work better than others. If the goal is to probe the magnitude of intermolecular electric fields, the frequency should be insensitive to field inhomogeneity and any intermolecular interaction other than electrostatics.
Our results indicate that H$_3$CNC and (CH$_3$)$_2$CO are the least sensitive to field inhomogeneity, while the results in Figure \ref{fig:fig2} show that the total frequency shift for (CH$_3$)$_2$CO most closely tracks the electrostatic contribution. We thus conclude that the CO stretch is optimal for reporting on intermolecular electric fields. We note this stretch was also found to be a better reporter of intermolecular fields than the CN stretch in a previous study\cite{fried2013measuring}. 

Of course, many stretching modes may be very useful reporters of intermolecular interactions even if they do not report on the electric field specifically. For example, numerous empirical relationships have been derived for the OH stretch frequency of water, enabling the prediction of vibrational spectra from simulation\cite{fecko2003ultrafast, corcelli2004combined, auer2007hydrogen, auer2008ir, choi2013computational,boyer2019beyond, lacour2023predicting} Likewise, the CN stretch is known to be an effective reporter of hydrogen bonding strength\cite{aschaffenburg2009probing,fafarman2010decomposition, baiz2020vibrational,Majumder2026}, enabling the measurement of hydrogen bonding strength under various conditions. Our work provides for a deeper understanding of why some of these relationships hold, and should enable the better use and interpretation of spectroscopic probes in the future.


\section*{\fontsize{12}{12}\selectfont 
SUPPORTING INFORMATION}
Additional results and analysis of vibrational frequencies.
Figure S1  shows how the probes respond to a point charge in both aligned and anti-aligned configurations.
Figure S2 shows  how the probes' equilibrium bonding length responds to point charges.
Figure S3 shows the anharmonic frequencies for the probes.
Figure S4 shows how the probes respond to electric fields generated by dipoles.
Figure S5 shows how the frequency of FH is influenced by the mass of its atomic constituents.

\section*{\fontsize{12}{12}\selectfont 
DATA AVAILABILITY}

The data that support the findings of this study are available within the article and its Supplementary Information. Additional relevant data and code are available from the corresponding authors upon request.

\section*{\fontsize{12}{12}\selectfont 
CODE AVAILABILITY} 
The in-house scripts used to generate all data in the manuscript will be made available from the corresponding author on reasonable request.

\section*{\fontsize{12}{12}\selectfont ACKNOWLEDGEMENTS}
\vspace{-3mm}
We thank the CPIMS program, Office of Science, Office of Basic Energy Sciences, Chemical Sciences Division of the U.S. Department of Energy under Contract DE-AC02-05CH11231. This work used computational resources provided by the National Energy Research Scientific Computing Center (NERSC), a U.S. Department of Energy Office of Science User Facility operated under Contract DE-AC02-05CH11231.

\bibliographystyle{achemso}

\bibliography{references}

\end{document}


\title{\textbf{Supplementary Information: The Interplay of Pauli Repulsion, Electrostatics, and Field Inhomogeneity for Blueshifting and Redshifting Vibrational Probe Molecules}}
\date{}
\author{R. Allen LaCour*$^{1,2}$, Ruoqi Zhao$^{1,2}$, Teresa Head-Gordon$^{1,2,3}$}
\maketitle
\noindent
\begin{center}
$^1$Kenneth S. Pitzer Theory Center and Department of Chemistry\\
$^2$Chemical Sciences Division, Lawrence Berkeley National Laboratory\\
$^3$Departments of Bioengineering and Chemical and Biomolecular Engineering\\
University of California, Berkeley, CA, 94720 USA

corresponding author: alacour@lbl.gov
\end{center}
\renewcommand{\theequation}{S\arabic{equation}}
\renewcommand{\thefigure}{S\arabic{figure}}
\renewcommand{\thetable}{S\arabic{table}}
\renewcommand{\bibnumfmt}[1]{(#1)}

\begin{figure}[H]
\centering
\includegraphics[width=0.9\textwidth]{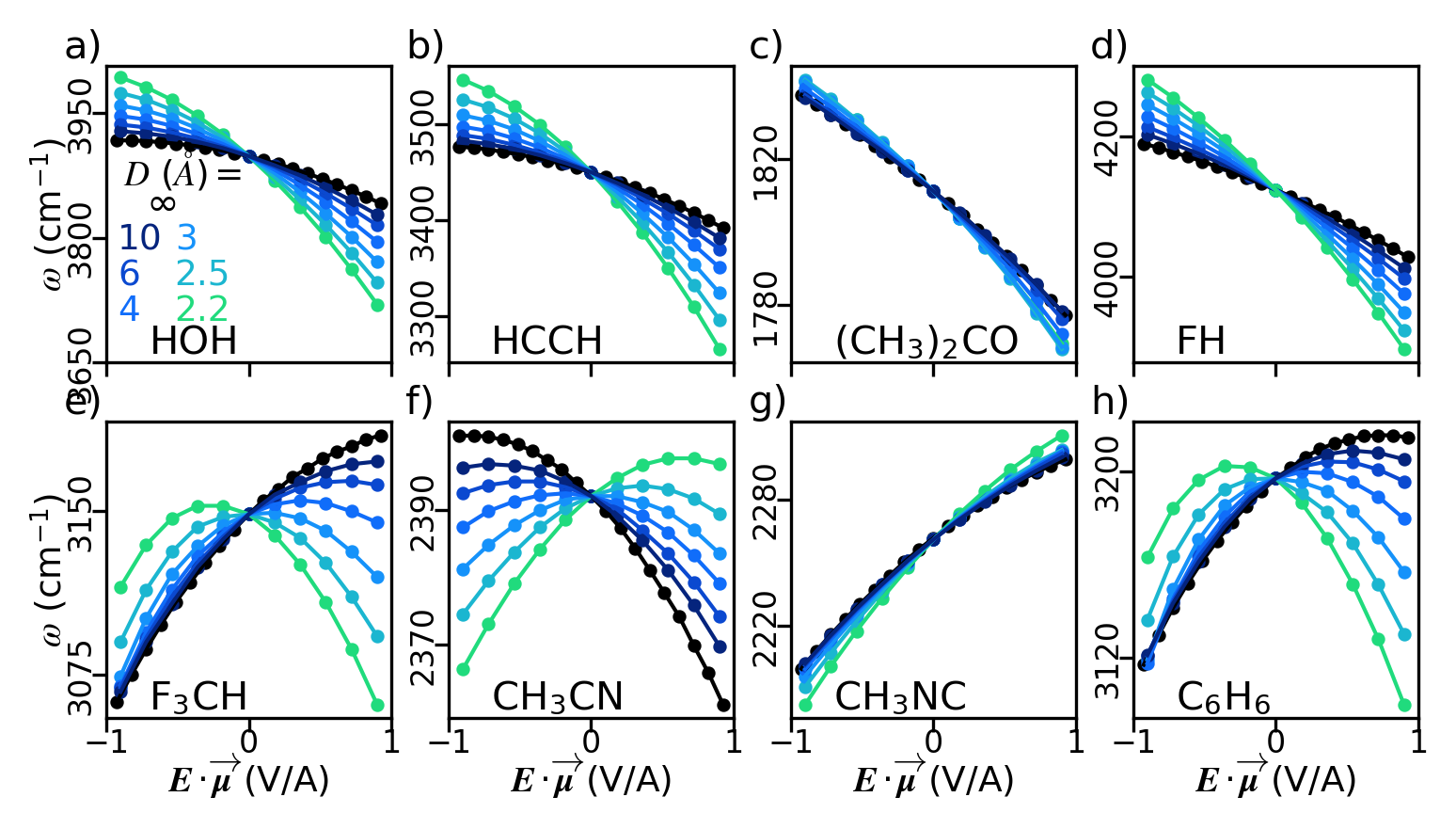}
\caption{\textbf{The frequency response of probes to the electric field generated by point charges with different electric field alignment.} Each panel shows the relationship between the stretching frequency ($\omega$) and the electric field ($\boldsymbol{E}$) experienced at the midpoint of the stretch. The vector $\boldsymbol{\overrightarrow{\mu}}$ is the direction of the bond dipole; thus positive or negative values of $\boldsymbol{E} \cdot \boldsymbol{\overrightarrow{\mu}}$ indicate that the field is aligned or anti-aligned with the bond dipole, respectively. The right-hand side (corresponding to alignment) was shown in Figure 1. The specific molecules examined are given in the bottom-left corner of each panel; the relevant stretching mode is between the last two atoms named. The legend in panel a) applies to all panels. Figure \ref{fig:fig1} in the main text. See Methods for how the frequencies are calculated.}
\label{fig:SIfig1}
\end{figure}

\begin{figure}[H]
\centering
\includegraphics[width=0.9\textwidth]{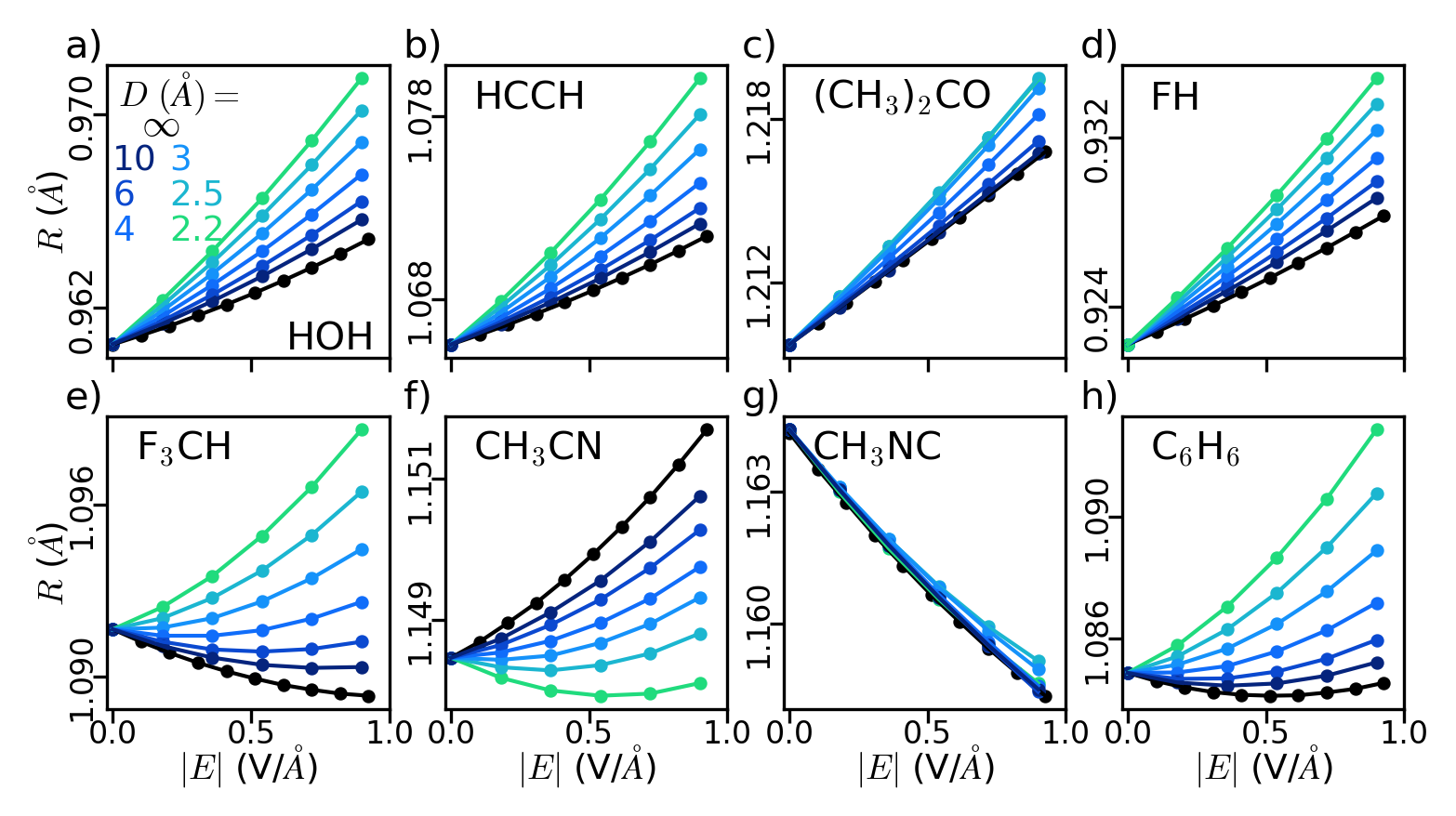}
\caption{\textbf{The change in equilibrium bond length due to the electric field generated by point charges.} The systems are identical to Figure 3.
Increased bond length is closely correlated with decreased frequency in Figure 3 and $vice$ $versa$ for decreased bond length.}
\label{fig:SIfig2}
\end{figure}

\begin{figure}[H]
\centering
\includegraphics[width=0.9\textwidth]{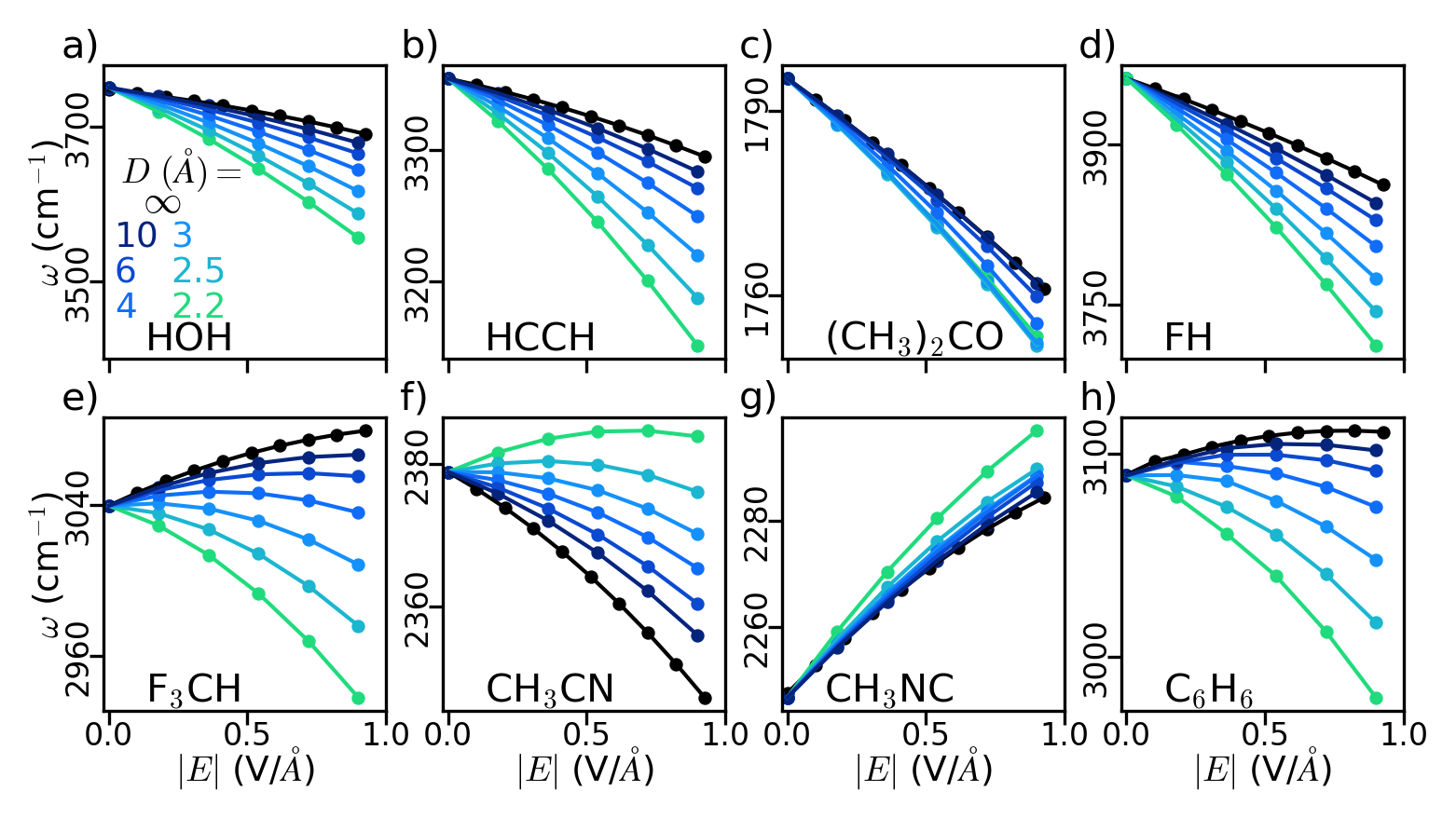}
\caption{\textbf{The anharmonic response of probes to the electric field generated by point charges.} The systems are identical to Figure 3 in the main text except that the anharmonic frequencies are used. 
Absolute values of the frequencies have changed, but the trends with increasing field inhomogeneity is identical to the harmonic frequencies.
See Methods for details on the calculation.}
\label{fig:SIfig3}
\end{figure}

\begin{figure}[H]
\centering
\includegraphics[width=0.95\textwidth]{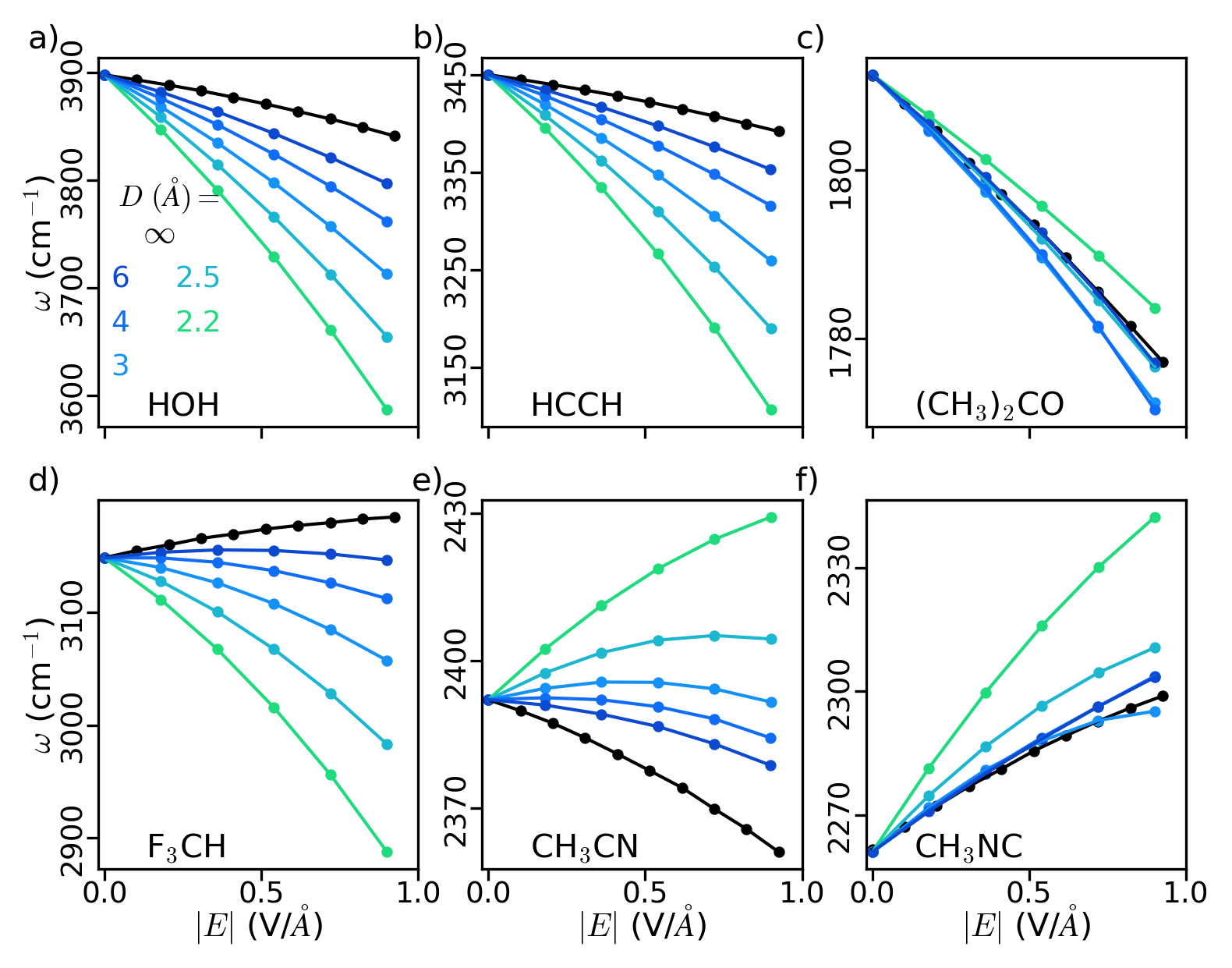}
\caption{\textbf{The frequency response of probes to the electric field generated by point dipoles.} Each panel shows the relationship between the stretching frequency ($\omega$) and the electric field ($\boldsymbol{E}$) experienced at the midpoint of the stretch. The sign of the point dipole was chosen based upon whether the probe prefers to interact with atoms carrying partial negative (a, b, d) or partial positive (c, e, f) charges.
Other details are the same as in Figure 3 in the main text.
Due to the more inhomogeneous field produced by point dipoles, the effect of inhomogeneity tends to be much larger in Figure S1 and S2. The sole exception is (CH$_3$)$_2$CO, which
is likely related to the non-linear angle at which it forms hydrogen bonds.
}
\label{fig:FigS4}
\end{figure}

\newpage

\begin{figure}[H]
\centering
\includegraphics[width=0.95\textwidth]{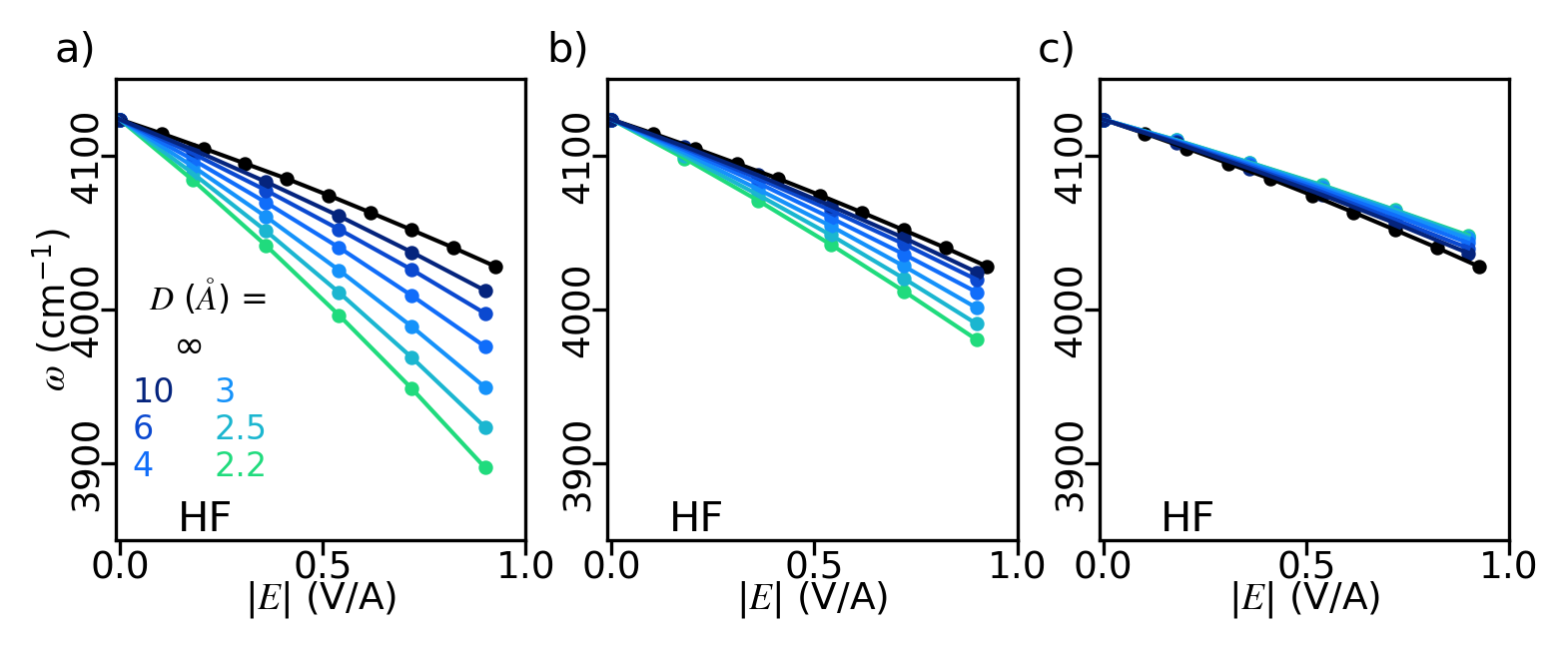}
\caption{\textbf{The influence of atomic mobility.} We repeat the calculations shown in Figure 3 for HF while varying the mass of the two atoms involved. In a), the actual masses of H and F are used, reproducing the data from Figure 2d.
In b), we set their masses to be equal, finding that the influence of inhomogeneity is dramatically reduced.
In c), we reverse their masses, finding that the inhomogeneity has almost no effect.
We kept the reduced mass of the stretch constant in all cases, so the mass only factors into the mobility of each atom during the stretch.
In other words, the sole difference between a) and c) is that in a) the H moves about 18 further than the F during the vibration while the opposite is true in c).
Our results support the idea that field inhomogeneity has a very large effect when the terminal atom is a hydrogen but is smaller when it is heavier.
}
\label{fig:FigS5}
\end{figure}